# Phase diagram and superconductivity at 58.1 K in α-FeAs free SmFeAsO$_{1-x}$F$_x$

M Fujioka[1], S J Denholme[1], T Ozaki[1], H Okazaki[1], K Deguchi[1,2], S Demura[1,2], H Hara[1,2], T Watanabe[1,2], H Takeya[1], T Yamaguchi[1], H Kumakura[1], and Y Takano[1,2]

[1] National Institute for Materials Science, 1-2-1 Sengen, Tsukuba, Ibaraki 305-0047, Japan

[2] University of Tsukuba, 1-1-1Tennodai, Tsukuba, Ibaraki 305-0001, Japan

E-mail: FUJIOKA.Masaya@nims.go.jp
TEL: +81 29-851-3354 (6321)
FAX: +81 29-859-2601

**Abstract**

The Phase diagram of SmFeAsO$_{1-x}$F$_x$ in terms of x is exhibited in this study. SmFeAsO$_{1-x}$F$_x$ from x = 0 to x = 0.3 were prepared by low temperature sintering with slow cooling. The low temperature sintering suppresses the formation of the amorphous FeAs, which is inevitably produced as an impurity by using high temperature sintering. Moreover, slow cooling is effective to obtain the high fluorine concentration. The compositional change from feedstock composition is quite small after this synthesis. We can reproducibly observe a record superconducting transition for an iron based superconductor at 58.1 K. This achievement of a high superconducting transition is due to the success in a large amount of fluorine substitution. A shrinking of the *a* lattice parameter caused by fluorine substitution is observed and the substitutional rate of fluorine changes at x =0.16.

## 1. Introduction

The first report of $SmFeAsO_{1-x}F_x$ was published in 2008 by Chen *et al* [1]. The superconducting transition temperature ($T_c$) was 43 K. Afterwards, $T_c$ was immediately improved to around 55 K by several groups [2-6]. Although, these groups adopted a high temperature sintering of around 1200 °C, some research for a low temperature sintering process had also been reported. For example, Chen *et al* synthesized polycrystalline $SmFeAsO_{1-x}F_x$ with a $T_c$ = 55 K by low temperature sintering at 1100 °C [7]. Wang *et al* shows superconductivity at 56.1 K, when the heating temperature is at 1000 °C [8] and Singh *et al* also reported superconductivity at 57.8 K, when the heating temperature is at 900 °C [9]. Therefore, low temperature sintering is very effective to obtain a sample with a high transition temperature. Moreover, it was found that $SmFeAsO_{1-x}F_x$ has a quite large upper critical magnetic field of over 100 T [10]. For this advantage, iron-based superconductors have been well studied as a promising material for applications as superconducting wires and films [11-14]. Ueda *et al* obtained $SmFeAsO_{1-x}F_x$ film with $T_c$ = 57.8 K by using molecular beam epitaxy [15]. They used the diffusion from an overlayer of $SmF_3$ to introduce the fluorine into the thin film.

The most important point for the synthesis of $SmFeAsO_{1-x}F_x$ is how to substitute fluorine for the oxygen site. In our previous study, it was found that fluorine is substituted not only at the maximum heating temperature but also during the cooling process. In particular, slow cooling is very effective to introduce much fluorine [16]. In this study, our samples are processed   It is also found that low temperature sintering, below the melting point of FeAs, does not form the amorphous FeAs impurity phase; which is inevitably produced by high temperature sintering of around 1200°C in the polycrystalline $SmFeAsO_{1-x}F_x$. Due to this impurity located between the superconducting grains of $SmFeAsO_{1-x}F_x$, the superconducting current has difficultly flowing over the grain boundaries. Therefore, low temperature sintering with slow cooling is very useful to obtain the amorphous FeAs-free $SmFeAsO_{1-x}F_x$ with high fluorine concentration [16].

In this study, we show a record high $T_c^{onset}$ (58.1 K) obtained by using this method. Moreover, we could obtain the high solid solubility limit of fluorine and produce a phase diagram of $SmFeAsO_{1-x}F_x$ in terms of x. While the phase diagram of $SmFeAsO_{1-x}F_x$ has been already reported by other two groups [2, 6], however, they prepared samples by using the high temperature sintering. This is the first report for the phase diagram of $SmFeAsO_{1-x}F_x$ obtained by using the low temperature sintering with slow cooling.

## 2. Experiment

$SmFeAsO_{1-x}F_x$ from x = 0 to x = 0.3 were prepared by using low temperature sintering with slow cooling, and x shows a nominal amount of fluorine concentration in this report. At first, two precursors namely 133 and 233 powder obtained by sintering the mixture of Sm, Fe and As metal powders at 850 °C for 10 h [13, 17]. Stoichiometric $Sm_2O_3$, $SmF_3$, 133 and 233 powders were mixed in a mortar and compressed into pellets. They were sintered at 980 °C for 40 h in evacuated quartz tubes, and cooled down at the rate of -5 °C /h to 600 °C.

X-ray diffraction (XRD; Rigaku Rint 2500) using Cu Kα radiation was measured for the characterization of the obtained samples. The lattice parameters and cell volumes were calculated from X- ray peak positions. Si powder (RIGAKU: RSRP-43275G) was used as an internal standard material. Their electrical resistivity was measured by the standard four-probe technique using Au electrodes. All samples were cut into almost the same size and sharp using the following dimensions (1.4 × 0.7 × 6.0 mm$^3$). In this study, $T_c^{onset}$ was regarded as the cross point of the fitting lines for resistivity in the normal state near transition and in the drop area during the transition. $T_c^{zero}$ was also regarded as the cross point of the line for zero resistivity and the $\rho$-$T$ curve. Magnetic measurements were performed with a SQUID magnetometer (Quantum Design: MPMS). The measurements were carried out under zero-field cooling (ZFC) and field cooling (FC).

## 3. Results and Discussion

Figure 1 shows a XRD pattern of SmFeAsO$_{0.84}$F$_{0.16}$. Black bars at the bottom show the calculated Bragg diffraction positions of SmFeAs(O,F). They correspond with the almost all of the obtained peaks except for Si peaks. However, several weak peaks from SmOF, SmAs and FeAs are also detected. Insert shows the expanded view near the main peak (102) of SmFeAs(O,F). Intensity of the peaks for SmOF and SmAs gradually increase when x > 0.18. On the other hand, in all samples with x < 0.16, their intensities are quite small and almost comparable. Additionally, the gradual shift of the main peak (102) to a higher angle is observed in the sample from x = 0 to x = 0.16. However, the shift is almost inhibited above x = 0.18. Lattice parameters and cell volume can be estimated by these peak shifts. The $a$ lattice parameter versus nominal amount of fluorine is shown in figure 2. From x = 0 to x = 0.04, it shows a gradually decreasing trend, and then, that slope rapidly becomes steep from x = 0.04 to x = 0.16. These decreases indicate an increase in fluorine concentration. When x > 0.16, although the reduction rate decreases, the $a$ lattice parameter linearly decreases with increasing nominal amount of fluorine. This is also suggested that fluorine continues to be slightly introduced into SmFeAs(O,F), when x > 0.16. When x = 0.26, the smallest value of the $a$ lattice parameter is observed. Inset shows cell volume versus nominal amount of fluorine. Similar behavior to the $a$ lattice parameter is observed.

Figure 3 shows a temperature dependence of resistivity for x = 0.02, 0.06, 0.08, 0.10, 0.16 and 0.26. The behavior of resistivity obviously changes according to the change of fluorine concentration. When x = 0.02, although superconductivity is not obtained, the anomalous kink is observed at around 145 K. This anomaly shifts to a lower temperature and the hump is smaller with increasing the fluorine concentration. When x = 0.08, the first appearance of $T_c^{onset}$ is observed at 7.6 K, however, resistivity does not reach to zero at 5 K. $T_c^{onset}$ rapidly increases with increasing fluorine concentration, and shows over 55 K, when x > 0.14. Moreover, the hump almost disappears, and the resistivity of normal state shows liner behavior. The expanded view of the temperature dependence of resistivity near $T_c$ is shown in figure 4. Although samples with 0.12 < x < 0.26 show the same behavior, $T_c^{onset}$ gradually increases. When x = 0.26, the highest value of $T_c^{onset}$ is observed at 58.1 K as shown in insert of figure 4. On the other hand, the maximum value of $T_c^{zero}$ is obtained at 53.7 K when x = 0.16. The decrease in $T_c^{zero}$ is observed in more than x = 0.16. $T_c^{onset}$, $T_c^{zero}$, $T_c^{anom}$ and the cell volume of each sample is listed in table 1.

The temperature dependence of magnetic susceptibility for the sample with x = 0.16 and 0.26 is shown in figure 5 and the demagnetization factor has been taken into consideration. These samples show a sharp diamagnetic transition and a high superconducting volume fraction. Inset shows ZFC and FC close to $T_c$. Although ZFC carve of the sample with x = 0.26 starts to rise at lower temperature compared with that of the sample where x = 0.16, higher $T_c$ is obtained in the sample with x = 0.26. The transition temperatures are observed at 57.5 K and 56.5 K. The $T_c$ obtained from magnetization measurement is slightly lower than those obtained from the resistivity measurement. Moreover, the superconducting volume fractions reach 97 % and 90 % in these samples (x = 0.16 and x = 0.26) respectively. The decrease in the superconducting volume fraction of the sample with higher fluorine concentration is thought to correlate with the increase in the impurity phases observed in figure 1.

Figure 6 displays the phase diagram of $SmFeAsO_{1-x}F_x$ in terms of x. The crystallographic transition from tetragonal to orthorhombic structure for undoped SmFeAsO is known at around 150 K [2, 6]. In this study, the kink of resistivity caused by this transition is observed at 147 – 144 K when x < 0.06. This transition disappears and superconductivity in $SmFeAsO_{1-x}F_x$ simultaneously appears between x = 0.06 and 0.08. $T_c^{onset}$ rapidly increases over x = 0.08. Whereas, the slope of the *a* lattice parameter changes at x = 0.16 as shown in figure 2, and the intensity of impurities start to increase from x = 0.16 as shown in the inset of figure 1. This means that the nominal amount of fluorine is almost same as the actual amount of fluorine, until x = 0.16. Moreover, it is also found that fluorine is not easy to be substituted over x = 0.16. However, the *a* lattice parameter gradually decreases over the nominal amount of x = 0.16. The maximum value of $T_c^{onset}$ is obtained at x = 0.26 and simultaneously this sample shows the smallest value of the *a* lattice parameter. $T_c^{onset}$ is decided by the strongest areas of superconductivity. Although the sample with x = 0.26 contains a lot of impurities, it also contains a large amount of fluorine. Therefore this sample locally composes a strong-superconducting area. On the other hand, $T_c^{zero}$ is decided by the weakest superconducting area in the polycrystalline $SmFeAsO_{1-x}F_x$. Therefore, when x > 0.16, because the samples include the many impurities, they form weak areas of superconductivity. Actually, gradual decreases in $T_c^{zero}$ are observed in the samples with x > 0.16. In this study, we enhanced the superconductivity up to 58.1 K. We might be able to obtain a higher $T_c$, if we succeed in futher fluorine substitution.

## 4. Conclusion

We obtained a record superconducting transition temperature for an iron based superconductor at 58.1 K by using low temperature sintering with slow cooling. This achievement of a high superconducting transition is due to the success in the large amount of fluorine substitution. The low temperature sintering with slow cooling does not form the amorphous FeAs phase in $SmFeAsO_{1-x}F_x$ and it is very effective for fluorine substitution. The substitutional rate of fluorine changes at x = 0.16. Then, the impurity phases start to increase from x = 0.16 and the superconducting volume fraction of this sample achieves 97 %. This means that, until x = 0.16, it is highly reliable in identifying the fluorine content in $SmFeAsO_{1-x}F_x$. We are able to make a phase diagram of $SmFeAsO_{1-x}F_x$ with more credibility.

**Acknowledgments**
This work was supported in part by the Japan Society for the Promotion of Science through 'Funding program

Table 1. Superconducting transition temperatures for each nominal amount of fluorine.

| x | $T_c^{onset}$ (K) | $T_c^{zero}$ (K) | $T^{anom}$ (K) |
|---|---|---|---|
| 0 | - | - | 147.1 |
| 0.02 | - | - | 145.2 |
| 0.04 | - | - | 144.7 |
| 0.06 | - | - | 129.5 |
| 0.08 | 7.6 | - | - |
| 0.1 | 39.1 | 30.31 | - |
| 0.12 | 54 | 49.7 | - |
| 0.14 | 55.9 | 51.3 | - |
| 0.16 | 57.4 | 53.7 | - |
| 0.18 | 57.6 | 53.5 | - |
| 0.2 | 57.7 | 53.3 | - |
| 0.22 | 57.8 | 52.9 | - |
| 0.24 | 58 | 52.7 | - |
| 0.26 | 58.1 | 52.2 | - |
| 0.28 | 57.7 | 50.9 | - |
| 0.3 | 57.9 | 51.2 | |

Figure 1. XRD patterns of SmFeAsO$_{0.84}$F$_{0.16}$ powders. Bottom bars indicate Bragg diffraction positions for SmFeAs(O,F). Insert shows an expanded view near main peak of SmFeAs(O,F).

Figure 2. The *a* lattice parameter versus nominal amount of x. Inset shows cell volume versus nominal amount of x.

Figure 3. Resistivity versus temperature from 5 K to 200 K for the sample with x = 0.02, 0.06, 0.08, 0.1, 0.16 and 0.26. Black and red arrows denote the anomalous kink and superconducting transition respectively.

Figure 4. Resistivity versus temperature for the samples from x = 0.12 to x = 0.26. Insert shows the expanded view near $T_c^{onset}$ of the sample with x = 0.26. Red lines are fitted lines for an estimation of $T_c^{onset}$.

Figure 5. Magnetic susceptibility versus temperature for the samples with x = 0.16 and x = 0.26. Inset shows the expanded view near $T_c^{onset}$ of the samples. Magnetic field is 1 Oe.

Figure 6. Phase diagram of SmFeAsO$_{1-x}$F$_x$ in terms of x. Green region denotes SmFeAs(O,F) with orthorhombic structure and the other region shows SmFeAs(O,F) with tetrahedral structure. Blue region denotes the area in which the superconductivity is obtained. When x > 0.16, SmFeAsO$_{1-x}$F$_x$ includes impurity phases.

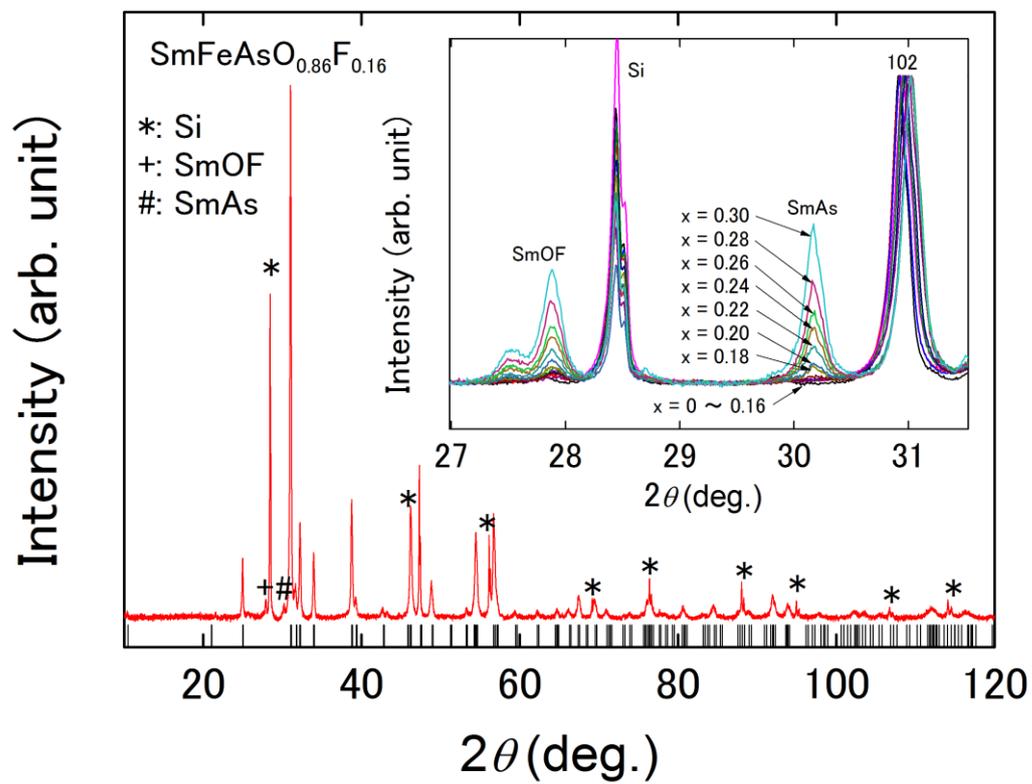

Fig. 1.

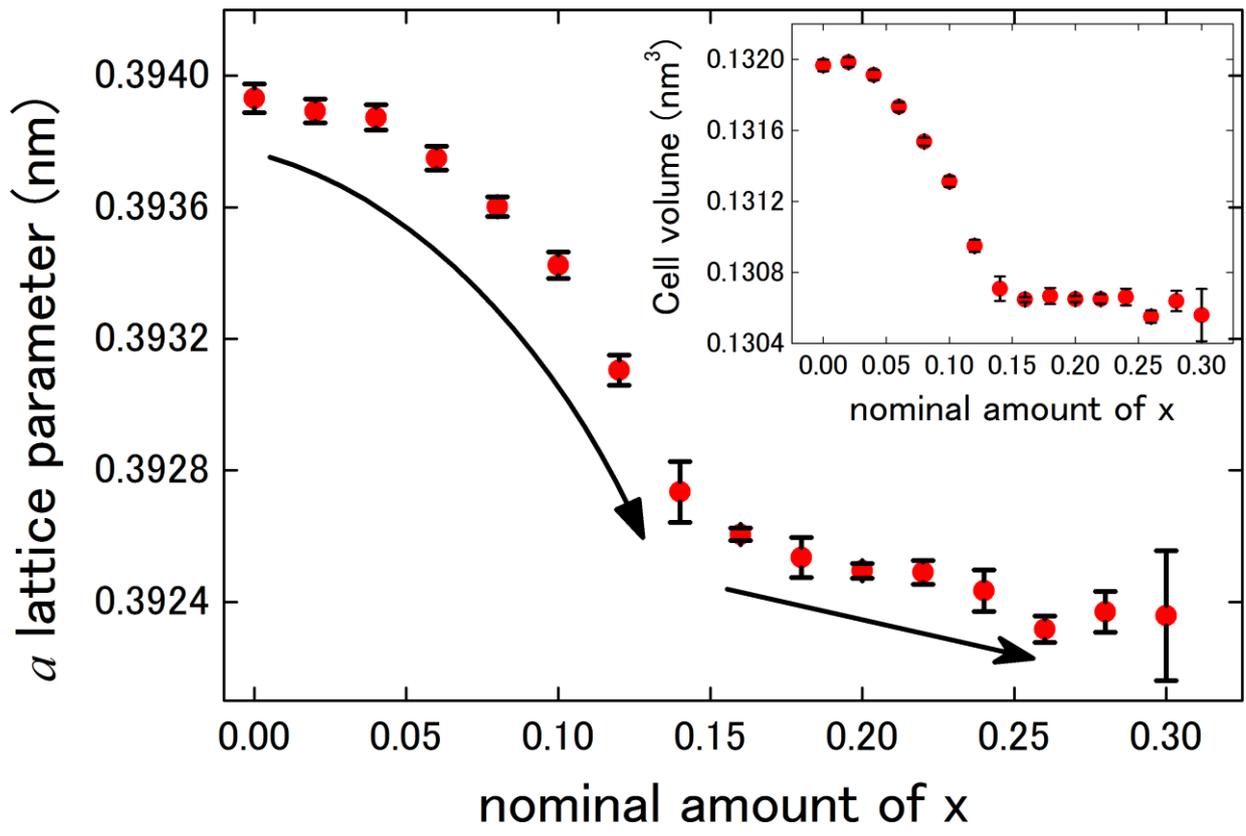

Fig. 2.

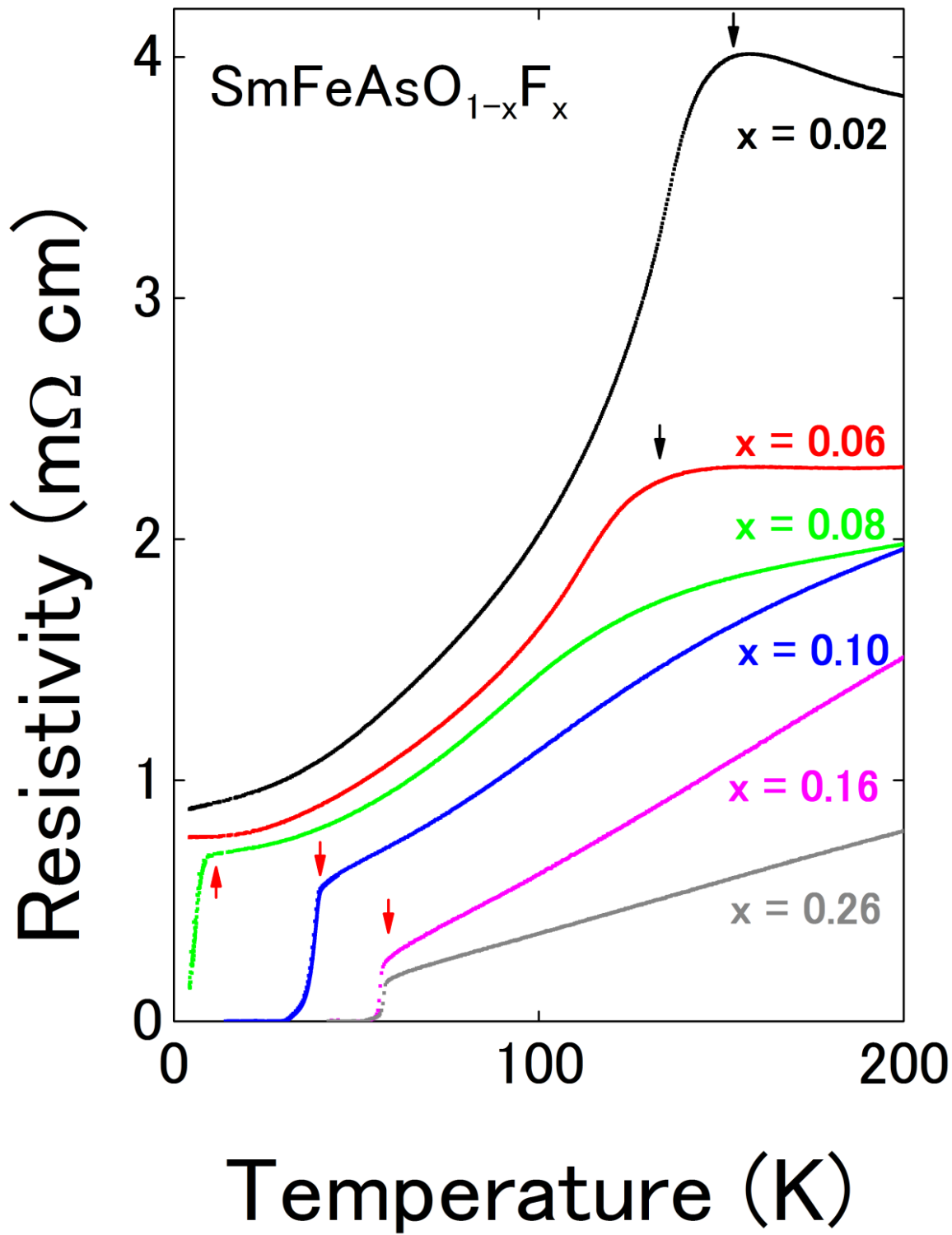

Fig. 3.

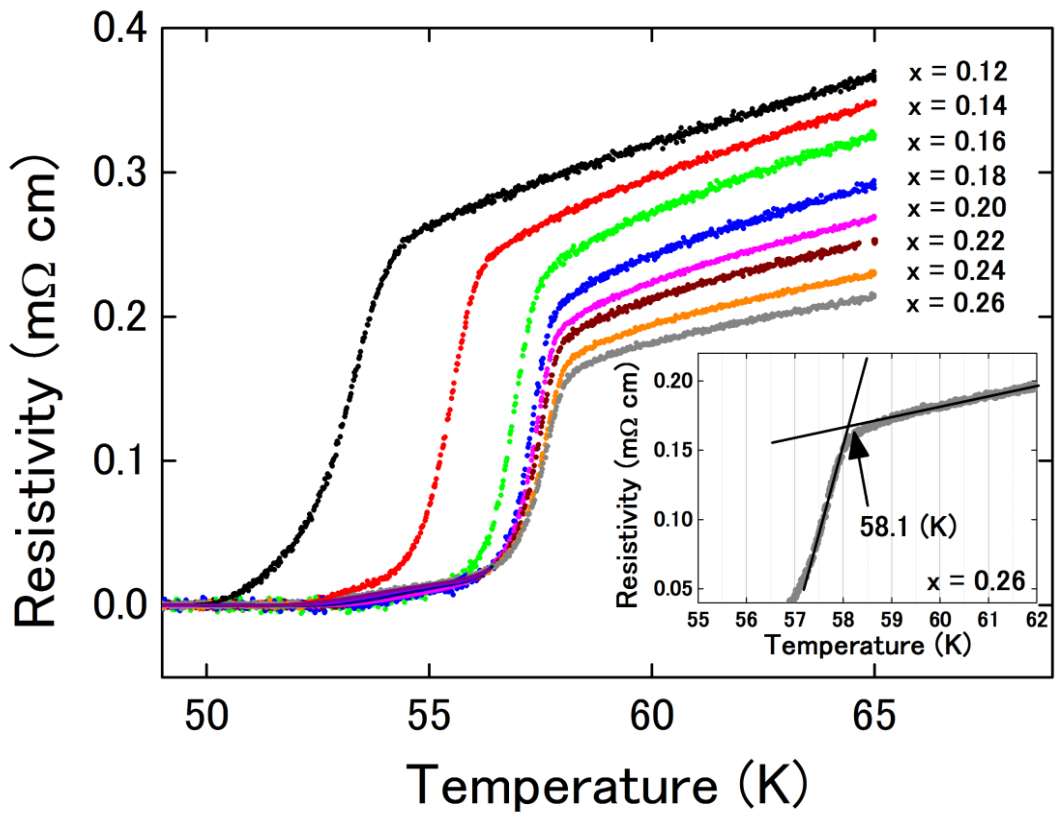

Fig. 4.

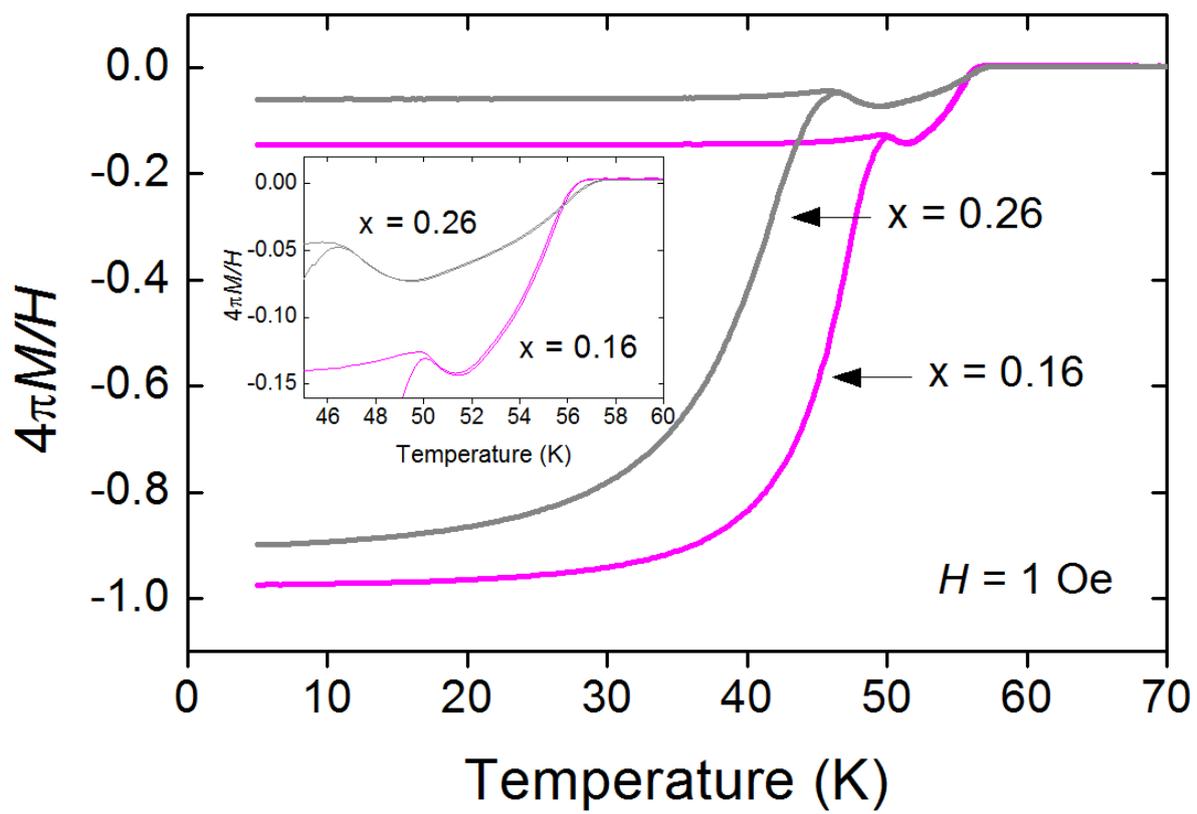

Fig. 5.

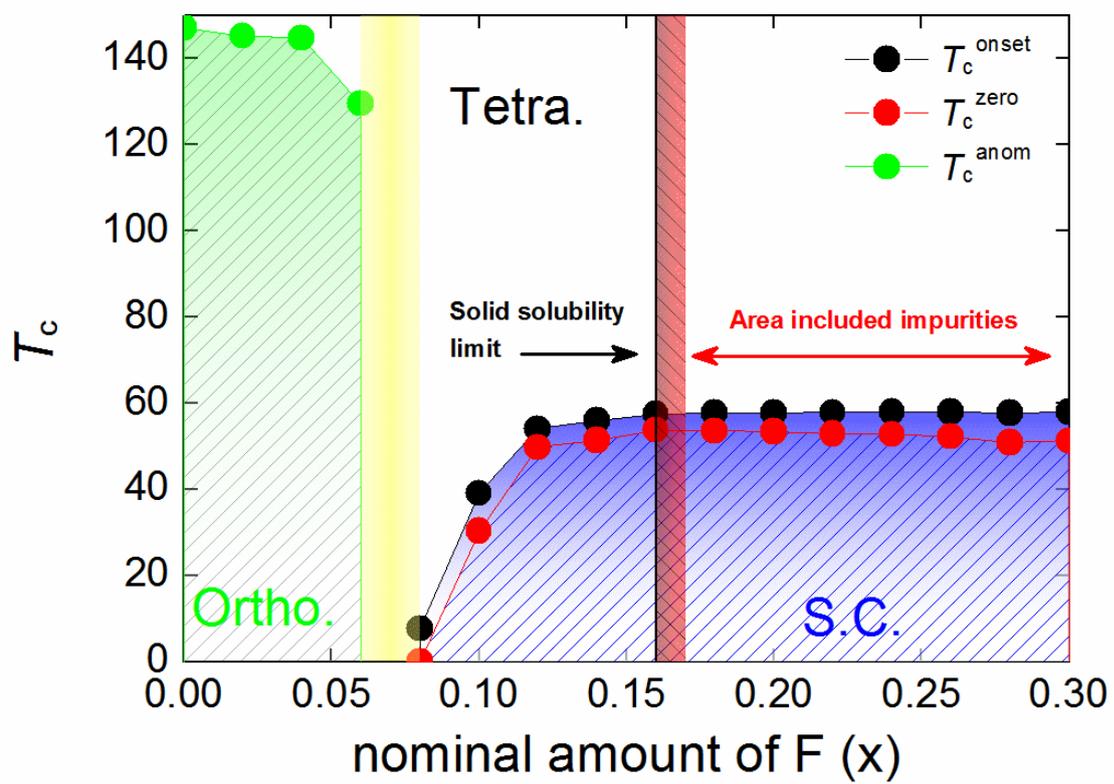

Fig. 6